\documentclass[english,11pt,oneside]{article}
\pdfoutput=1
\usepackage{amsmath}
\usepackage{amsfonts}
\usepackage{amssymb}
\usepackage{dsfont}
\usepackage{graphicx, rotating}
\usepackage{epstopdf}
\usepackage{epsfig}
\usepackage{latexsym}
\usepackage{multirow}
\usepackage{color}
\usepackage{slashed}
\usepackage{pifont}
\usepackage[dvipsnames]{xcolor}
\usepackage{mathtools}
\usepackage[top=2.8 cm, bottom=2.8 cm, left=3 cm, right=3 cm]{geometry}
\usepackage[compat=1.0.0]{tikz-feynman}
\usetikzlibrary{shapes.misc}
\tikzset{cross/.style={cross out, draw=black, minimum size=2*(#1-\pgflinewidth), inner sep=0pt, outer sep=0pt},
	cross/.default={1pt}}
\usepackage{subcaption}
\newcommand{\ba}{\begin{eqnarray}}
\newcommand{\ea}{\end{eqnarray}}
\newcommand{\no}{\nonumber}

\newcommand\blfootnote[1]{%
	\bgroup
	\renewcommand\thefootnote{\fnsymbol{footnote}}%
	\renewcommand\thempfootnote{\fnsymbol{mpfootnote}}%
	\footnotetext[0]{#1}%
	\egroup
}

\usepackage{hyperref}
\hypersetup{
 colorlinks = true,
 linkcolor  = red,
 citecolor={blue}
}

\usepackage{mfirstuc}
\newcommand{\addReviewer}[2]{
	\expandafter\newcommand\csname #1\endcsname[1]{{\textbf{ \color{#2} \capitalisewords{#1}:\,##1}}}
	\expandafter\newcommand\csname #1cor\endcsname[2]{{\color{#2} \capitalisewords{#1}:\,\st{##1}{\textbf{##2}}}}
	\expandafter\newcommand\csname #1color\endcsname{#2}
	\expandafter\newcommand\csname #1todo\endcsname[1]{{\todo[inline,color=white!70!#2, caption={}]{\textbf{\capitalisewords{#1}}: ##1}}}
}
\addReviewer{AV}{Salmon}

\usepackage{bbold}

\usepackage[nottoc]{tocbibind}

\numberwithin{equation}{section}

\begin{document}

\unitlength = 1mm

\thispagestyle{empty} 
\begin{center}
	\vskip 3.4cm\par
	{\par\centering \textbf{\LARGE Vector-like symmetries and parity conservation\\[0.25cm] in gauge theories with Yukawa couplings}}
	
	\vskip 1.2cm\par
	{\scalebox{.85}{\par\centering \large  
			\sc\hyperlink{AV}{\color{black}Alessandro Valenti}$\,^{a,b}$, \hyperlink{LV}{\color{black}Luca Vecchi}$\,^{b}$}
		{\par\centering \vskip 0.7 cm\par}
		{\sl 
			$^a$~Dipartamento di Fisica e Astronomia ``G.~Galilei", Università di Padova, Italy
		}\\
		{\par\centering \vskip 0.2 cm\par}
		{\sl 
			$^b$~Istituto Nazionale di Fisica Nucleare, Sezione di Padova, I-35131 Padova, Italy
		}\\
		
		{\vskip 1.65cm\par}}
\end{center}

\begin{abstract}

Non-perturbative results in QCD-like theories can be derived employing positivity of the Euclidean path integral measure, as pioneered by Weingarten, Vafa, Witten. We show that positivity of the measure can be generalized to parity-invariant theories with Yukawa couplings to fundamental scalars, provided the fermions are Dirac and carry a real representation of the gauge group. This result allows us to demonstrate the conservation of parity and vector-like flavor symmetries in such theories, as well as to derive exact inequalities among hadrons' masses.

\end{abstract}

\blfootnote{
	\hypertarget{AV}{\href{mailto:alessandro.valenti@pd.infn.it}{\color{black}{alessandro.valenti@pd.infn.it}}}
}
\blfootnote{
	\hypertarget{LV}{\href{mailto:luca.vecchi@pd.infn.it}{\color{black}{luca.vecchi@pd.infn.it}}}
}

\newpage

{
	\hypersetup{linkcolor=black}
	\tableofcontents
}

\section{Introduction}

Non-perturbative results in non-supersymmetric four-dimensional gauge theories are very rare. Among the few examples that come to mind are the inequalities in massive QCD-like scenarios established in the seminal work of Weingarten \cite{Weingarten:1983uj} and Vafa-Witten \cite{Vafa:1983tf,Vafa:1984xg}. Central to these results is the {\emph{positivity}} of the Euclidean path integral measure, derived in \cite{Vafa:1983tf} assuming a positive definite fermion mass and a vanishing topological angle. With these conditions at hand, inequalities between functionals of the bosonic fields can be turned into correlator inequalities and eventually employed to prove relations among hadron masses, as well as the absence of spontaneous symmetry breaking of vector-like flavor symmetries and parity. (See \cite{Nussinov:1999sx} for a review and independent supporting arguments and \cite{Einhorn:2002rm} for a discussion of P-conservation). 

Unfortunately, those results cannot be straightforwardly extended to chiral gauge theories because in such cases the fermionic determinant is usually complex-valued. A generalization to scenarios with Yukawa couplings to fundamental scalars is also not possible in general. Parity-even scalars generically lead to non-positive effective masses for the fermions and thus spoil positivity, whereas pseudo-scalar couplings even violate reality of the fermionic determinant. An obvious way for recovering positivity in QCD-like theories with Yukawa couplings to P-even scalars is via a doubling of the fermionic fields, similarly to what was done in \cite{Weingarten:1983uj}; this way the full fermionic contribution to the path integral becomes the square of a real quantity and is thus positive. A different flavor symmetry has been proposed in ref. \cite{Vafa:1983tf} to recover strict positivity in scenarios with pseudo-scalars.

With the discovery of the Higgs boson it became apparent that fundamental scalars are not only hypothetical particles motivated by Supersymmetry and phenomenological scenarios beyond the Standard Model, but in fact key players in Nature. It is therefore not unconceivable that one day a strongly-coupled scalar will be found to have actual physical relevance. Asymptotically-free gauge theories with scalars represent very interesting QFTs as they are anticipated to feature a rather non-trivial IR physics due to the rich coupling structure, which includes potentially {\emph{relevant}} Yukawa interactions as well as scalar self-interactions. And yet, very little is known about their dynamics at the non-perturbative level. It is therefore important to see whether more can be learned about them.

In this paper we present a large class of non-Supersymmetric asymptotically-free gauge theories with Yukawa couplings for which exact results can be established. Our scenarios require Dirac fermions in {\emph{real}} representations of the gauge group as well as parity invariance. Other than baryon number, no additional flavor symmetries are invoked. We begin in Section \ref{sec:posi} by proving positivity of the Euclidean path integral in theories with one Dirac fermion and neutral P-even scalars, and then generalize the result to multiple flavors and (P-even or P-odd) scalars with gauge quantum numbers. Subsequently, in Section \ref{sec:appl} we demonstrate the conservation of parity and vector-like flavor symmetries (defined as the fermion symmetries respected by the masses and Yukawa couplings), and finally discuss hadron mass inequalities.

\section{Positivity of the measure}
\label{sec:posi}

Consider a four-dimensional gauge theory with two Weyl fermions $\psi_\pm$ charged under the same real (as opposed to pseudo-real) representation ${\cal R}$ of the gauge group. Our main focus is asymptotically-free theories, which have a non-trivial IR dynamics. Let us then add a real gauge-singlet scalar $\phi$ with Yukawa couplings to $\psi_\pm$. We further demand invariance under parity, with $\phi$ assumed to be P-even, and $\psi_\pm\to e^{\pm i\beta}\psi_\pm$, a $U(1)_B$ subgroup of the approximate non-anomalous $SU(2)$ flavor symmetry which we call {\emph{baryon number}}.~\footnote{Note that in our scenarios also bosonic hadrons can carry this quantum number.} The theory is automatically invariant under charge conjugation as well. Generalizations of this setup will be presented in Section \ref{sec:gene}. 

Within our assumptions we can embed the fermions in a single Dirac field $\Psi=(\psi_+,\eta\psi_-^\dagger)$, with $\eta$ the unitary invariant tensor of ${\cal R}$. After a phase re-definition we can write the fermionic part of the Euclidean action as
\ba\label{pseudoL}
S^{\rm fermions}_E=\int {\rm d}^4x~\overline{\Psi}[{\slashed{D}}+M_\phi]\Psi,
\ea
where ${\slashed{D}}=\gamma^\mu D_\mu$, $\left\{\gamma^\mu,\gamma^\nu\right\}=2\delta^{\mu\nu}$, $D_\mu=\partial_\mu-iA_\mu$, $A^\dagger_\mu=A_\mu$, and 
\ba\label{Mphi}
M_\phi=m+y\phi 
\ea
is the sum of a real mass and a real Yukawa coupling. The $\gamma^\mu$ matrices are taken to be hermitian throughout the paper. 

The fermions can be integrated out exactly, with a formal result given by 
\ba\label{detDM}
\int{\cal D}[\Psi,\overline\Psi]~e^{-S^{\rm fermions}_E}={\rm det}[{\slashed{D}}+M_\phi]. 
\ea
Throughout the paper we implicitly adopt a regularization that preserves the key relation \eqref{realR} (see below), as well as $\gamma_5$-hermiticity (${\slashed{D}}^\dagger=\gamma_5{\slashed{D}}\gamma_5$). An example is provided by a finite lattice with Wilson fermions or Pauli-Villars. Moreover, our results are derived for an arbitrarily large but {\emph{finite}} value of the UV regulator $1/\ell$ in order to avoid a triviality problem associated to the presence of the fundamental scalar. As long as the bare couplings satisfy $y^2\ll g^2\ll\lambda_\phi\ll16\pi^2$, with $\lambda_\phi>0$ denoting the quartic $\phi$ self-coupling(s) and $g$ the gauge coupling, the interactions stay perturbative down to an exponentially smaller confinement scale $\Lambda$ of the non-abelian dynamics. At that scale, where all the interesting dynamics unfolds, the effect of the UV completion is strongly suppressed by powers of $\Lambda\ell$. Hence, while we are not able in general to take a strict continuum limit, the existence of a large (finite) UV cutoff is not expected to qualitatively impact our conclusions.~\footnote{The same assumption is necessary in order to extend the validity of the results of \cite{Vafa:1983tf} to real-world QCD, where the cutoff is set by the $W^\pm$ mass.}

Using $\gamma_5$-hermiticity follows that the functional determinant \eqref{detDM} is real for any representation of the gauge group, provided $M_\phi$ is real (which here follows from P). For fermions in a real representation one can go one step further and prove that it is also non-negative:
\ba\label{theorem}
{\rm det}[{\slashed{D}}+M_\phi]\geq0.
\ea
To proceed we first put our system in a finite box of large 4-volume $V_4$ such that the spectrum becomes discrete. The operator ${\slashed{D}}+M_\phi$ is however not normal because $M_\phi$ is a function. Rather than calculating the determinant of ${\slashed{D}}+M_\phi$ directly, we find it convenient to introduce the definition $H\equiv\gamma_5({\slashed{D}}+M_\phi)$. Such operator is hermitian and therefore diagonalizable, with ${\rm det}[H]$ given by the product of its eigenvalues. We will next show that the determinant of $H$ is positive and subsequently that ${\rm det}[H]={\rm det}[{\slashed{D}}+M_\phi]$.  

We will now prove that the determinant of $H$ is positive by showing that its (real) eigenvalues have even multiplicity whenever the fermions are in real representations of the gauge group. In four-dimensional Euclidean space, and recalling that our gamma matrices are hermitian, there exists a unitary real matrix ${\cal T}$ satisfying ${\cal T}\gamma^*_\mu {\cal T}^{\dagger}=\gamma_\mu$ and ${\cal T}^t=-{\cal T}$. 
Furthermore, for any self-conjugate representation $\eta D_\mu^*\eta^\dagger= D_\mu$. Because $M_\phi$ is a gauge-invariant and $SO(4)$-singlet real function, it follows that
\ba\label{realR}
(\eta {\cal T})H^*(\eta {\cal T})^{\dagger}=H.
\ea
Denote next by $\chi_a$ a complete set of linearly independent c-valued eigenvectors of $H$ associated to an eigenvalue $\lambda$, with the index $a=1,\cdots,n_\lambda$ running through the dimensionality $n_\lambda$ of the kernel of $(H-\lambda)$. According to \eqref{realR}, to any $\chi_a$ there is a corresponding $\eta {\cal T}\chi^*_a$ with the same real eigenvalue. Because by assumption our set $\chi_a$ is complete, the eigenvectors must satisfy a reality condition
\ba\label{Nf=1}
\eta {\cal T}\chi^*_{a}=c_{ab}\chi_{b}
\ea
for some matrix $c_{ab}$. When considering {\emph{real representations}}, $\eta$ is symmetric and $(\eta {\cal T})(\eta {\cal T})^*=-1$. Multiplying \eqref{Nf=1} on the left by $(\eta {\cal T})^*$ and using the complex conjugate version of \eqref{Nf=1} we obtain $cc^*=-1$. Taking the determinant of the left and right hand side it is easy to see that the relation $cc^*=-1$ can only be satisfied if the multiplicity $n_\lambda$ is even. Each eigenvalue contributes to ${\rm det}[H]$ via the combination $\lambda^{n_\lambda}$, and so ${\rm det}[H]\geq0$, as we wanted to prove. Because this is valid for any $V_4$, it also holds for $V_4\to\infty$.

The pairing of the eigenvalues of $H$ may also be understood as a consequence of a generalization of Kramers' theorem of Quantum Mechanics. To see how this works observe that eq. \eqref{realR} is equivalent to the statement that the ``Hamiltonian" $H$ commutes with the anti-unitary ``time reversal" operator ${\mathsf T}$ defined as $\chi\to{\mathsf T}\chi=\eta {\cal T}\chi^*$.~\footnote{We use quotation marks because this is not time reversal, for example it does not act on the coordinates nor does it rotate the gauge fields.} Crucially, for fermions in real representations of the gauge group (as opposed to pseudo-real) ${\mathsf T}^2=-1$. As a consequence, for each (c-valued) eigenvector $\chi$ there is always a linearly independent eigenvector ${\mathsf T}\chi$ with the same eigenvalue, in agreement with what argued below \eqref{Nf=1}.

In order to establish \eqref{theorem} it remains to demonstrate that ${\rm det}[H]={\rm det}[{\slashed{D}}+M_\phi]$. To see this it is enough to show that ${\rm det}[\gamma_5]=1$. It is convenient to work within a basis of eigenstates of the anti-hermitian operator ${\slashed{D}}$. We call $u_{\mu}$ the complete set of eigenvectors of ${\slashed{D}}$ with eigenvalue $i\mu$ ($\mu\in{\mathbb R}$, including zero). Analogously to \eqref{realR} we have $(\eta {\cal T}){\slashed{D}}^*(\eta {\cal T})^{\dagger}={\slashed{D}}$, which reveals that $\eta {\cal T}u^*_{\mu}=c'\gamma_5u_{\mu}$ for some even-dimensional matrix $c'$ satisfying $c'c'^*=-1$. Hence the eigenvalue $\mu$ has even degeneracy. The number of $\mu>0$ is the same as the number of $\mu<0$ because $\left\{{\slashed{D}},\gamma_5\right\}=0$ and without loss of generality we can define $u_{-\mu}=\gamma_5u_\mu$. Zero-modes can be chosen to be simultaneously eigenstates of $\gamma_5$, and our by now familiar ``reality" condition implies that the number $n_+$ and $n_-$ of positive and negative chirality zero-modes are separately even. Within the basis $\left\{u_{\mu>0},\gamma_5u_{\mu>0},u_0^\pm\right\}$ one explicitly finds ${\rm det}[\gamma_5]=(-1)^{N_\pm}(-1)^{n_-}=1$, where $N_-=N_+$ is the number of non-vanishing negative or positive eigenvalues. Both $N_\pm$ and $n_-$ are even and so ${\rm det}[\gamma_5]=1$. This completes the proof of \eqref{theorem}.

A few comments are in order:

i) The positivity of the path integral for fermions in real representations and {\emph{constant mass}} has been noted before, see \cite{Montvay:1997ak,Hands:2000ei} for Dirac and \cite{Cherman:2019hbq,Karasik:2022gve} for Weyl fermions. What we demonstrated above is that positivity applies to P-conserving Dirac theories with Yukawa couplings to P-even scalars, i.e. to any real function $M_\phi$. We will see in the following that this result can be generalized to theories with charged scalars as well as multiple flavors. This will allow us to analytically establish non-perturbative results in such theories for the first time.

ii) Unlike QCD, in our models the two Weyl fermions $\psi_\pm$ transform as the same real representation. And yet it should be noted that our positivity theorem \eqref{theorem} is not a trivial consequence of a doubling of the fermionic field degrees of freedom. Let us see why. Performing a non-anomalous change of variables $(\psi_+,\psi_-)\to(\psi_1,\psi_2)$ the mass matrix becomes the identity in flavor space times $iM_\phi$. The baryon number now acts as an ordinary $SO(2)_B$ rotation on $(\psi_1,\psi_2)$ and in fact the entire classical action enjoys a manifest $SO(2)_B$ symmetry. Naively, the fermionic path integral appears to be the square of the (complex) Pfaffian ${\rm Pf}[C({\slashed{D}}+i\gamma_5M_\phi)]$. That would however lead to an overall complex result, which contradicts our positivity theorem. In fact, one should also take into account a (topological sector-dependent) proportionality factor coming from the re-ordering of the grassmanian integration variables emerging from the previous change of variables \cite{Stone:2020vva}. The $SO(2)_B$ flavor symmetry of the Lagrangian is therefore not enough to claim that the path integral is trivially the square of a real quantity.

iii) As a corollary of our theorem we observe that the overall sign of $M_\phi$ is unphysical. In fact, the chiral rotation needed to $M_\phi\to-M_\phi$ shifts the theta angle by $2\pi T_{\cal R}$, with $T_{\cal R}$ the Dynkin index of the representation ${\cal R}$. The index theorem relates the difference between zero-modes of ${\slashed{D}}$ with positive chirality ($n_+$) and those with negative chirality ($n_-$) to the topological number $\nu=({g^2}/{32\pi^2})\int {\rm d}^4x\, F^A_{\mu\nu}\widetilde F^{A\,\mu\nu}$ via 
\ba\label{indexTH}
n_+-n_-=2T_{\cal R}\nu. 
\ea
The topological angle appears in the path integral via the combination $e^{i\theta\nu}$. Hence, the change in the sign of the fermion mass introduces a trivial phase $e^{i2\pi T_{\cal R}\nu}=e^{i(n_+-n_-)\pi}=1$ since we have seen that $n_\pm$ are even. When the mass is constant, without loss of generality we can thus take $M_\phi$ positive and our positivity theorem reduces to a triviality. The point is of course that we are ultimately interested in scenarios in which $M_\phi$ is a field. In those cases positivity is still not obvious but the general proof given above guarantees it.

\subsection{Generalizations}
\label{sec:gene}

\subsubsection*{Charged (pseudo-)scalars}
\label{sec:gencharged}

In deriving \eqref{theorem} we assumed $\phi$ was P-even and gauge-singlet. We now want to explore scenarios with different P and gauge charges.

Our theorem readily generalizes to scenarios where the scalars are charged under the gauge group provided the condition $\eta M_\phi^*\eta^\dagger=M_\phi$ remains valid and fermions, as always, are in real representations ($\eta^t=\eta$). Yet, we find that the P-charge assignments of $\phi$ necessary to ensure the condition $M_\phi^*=M_\phi$ depend on the gauge quantum numbers. A few concrete examples will better illustrate this point. Alternative theories can be constructed along these lines.

Let us assume for definiteness the fermions are in the fundamental representation of an $SO(N_c)$ gauge group. The scalar can then be in the trivial representation, the 2-index symmetric traceless ($\phi_{(AB)}$) or the anti-symmetric ($\phi_{[AB]}$) representation ($A,B$ belong to the fundamental of $SO(N_c)$). The trivial representation has been considered in Section \ref{sec:posi} assuming $\phi$ was P-even. If $\phi$ was a pseudo-scalar gauge-singlet we would have to replace $y\phi$ with $i\gamma_5y\phi$ in our proof, where the $\gamma_5$ is necessary to preserve P whereas the complex $i$ ensures that the (Lorentzian signature) action be hermitian. Unfortunately in such a case reality and positivity cannot be proven since the relation $M_\phi^*=M_\phi$ is violated. Hence, our theorem applies to scalar singlets only if they are P-even.

The coupling to a (real) P-even scalar in the symmetric representation may be written as 
\ba
y\overline{\Psi}_{A}\Psi_{B}\phi_{(AM)}\eta^*_{MB}.
\ea
The scalar is real in the sense that $\phi^*=\eta^\dagger\phi\eta^*$. The fermionic Lagrangian can again be written as in \eqref{pseudoL} with $[M_\phi]_{AB}=\delta_{AB}m+y\phi_{(AM)}\eta^*_{MB}$ hermitian and satisfying the desired relation $\eta[M_\phi]^*\eta^\dagger=[M_\phi]$. Positivity of the fermionic determinant can be proven exactly as for the scalar singlet. This conclusion would however be invalidated if $\phi_{(AB)}$ was P-odd, similarly to the scalar singlet scenarios.

The situation is somewhat reversed if the real $\phi_{[AB]}$ is in the anti-symmetric representation, though. In such a case to ensure positivity we must require $\phi_{[AB]}$ be P-odd, so the Yukawa interaction
\ba\label{aps}
y\overline{\Psi}_{A}\gamma_5\Psi_{B}\phi_{[AM]}\eta_{MB}^*,
\ea
has a real $y$, as can be seen enforcing hermiticity of the action in Minkowski space. That way we have the same formal structure as in \eqref{pseudoL}, but now with $[M_\phi]_{AB}=\delta_{AB}m+y\phi_{[AM]}\eta_{MB}^*\gamma_5$. The condition \eqref{realR} is satisfied and our theorem goes through unchanged. As opposed to before, if $\phi_{[AB]}$ was taken to be P-even the coupling would be purely imaginary and reality of the path integral, along with its positivity, would be lost.

We see that, once a set of gauge representations for the fermions and scalars are chosen, the parity properties of $\phi$ are determined by the requirement $\eta M_\phi^*\eta^\dagger=M_\phi$. For this basic reason, our theorem cannot be extended to Supersymmetric gauge theories, where the scalars necessarily come in P-even and P-odd pairs.

\subsubsection*{Many flavors}
\label{sec:geneNf}

In the presence of $N_f$ Dirac flavors coupled to real P-even scalar singlets $\phi$ the quantity $M_\phi$ in \eqref{Mphi} becomes a hermitian matrix $[M_\phi]_{ij}=m_{ij}+y_{ij}\phi$ in flavor space. Our earlier positivity theorem  \eqref{theorem} can then be extended to this case provided $M_\phi$ is {\emph{real}}: for any real matrix $M_\phi$ the relation \eqref{realR} says that the eigenvalues of $\gamma_5({\slashed{D}}+M_\phi)$ appear in pairs. Thus, positivity of the fermionic determinant can be established for an arbitrary number $N_f$ of Dirac fermions in real representations of the gauge group as long as $M_\phi=M_\phi^*$. That requirement can for example be naturally ensured imposing charge-conjugation. This in turn demands appropriate C-charges for the scalars; for example, the P-even singlet $\phi$ must also be C-even. Scenarios with charged scalars can be treated analogously. (The necessary C-charges in general depend on the gauge representation.)

\subsubsection*{Pseudo-real representations?}

Our proof of the even degeneracy of the $H$ eigenvalues cannot be generalized to Yukawa theories of a single Dirac fermion in a pseudo-real representation. Technically, this is because when considering pseudo-real representations the invariant tensor $\eta$ is antisymmetric, so the matrix $c$ entering the reality condition corresponding to \eqref{Nf=1} satisfies $cc^*=1$ and is allowed in principle to have any dimensionality. {Equivalently, for pseudo-real representations ${\mathsf T}^2=+1$.}

Nevertheless, one may invoke flavor symmetries to enforce positivity. An obvious way to achieve this is to impose a $Z_4\subset SO(2)$ that exchanges pairs $(\Psi_{1},\Psi_{2})$ of Dirac fermions according to $(\Psi_{1},\Psi_{2})\to(\mp\Psi_{2},\pm\Psi_{1})$ and hence effectively doubles the fermionic fields, implying that ${\rm det}[H]$ is the square of a real quantity. There also seems to be a superficially different, more sophisticated possibility, though. Indeed, consider a theory with an even number $N_f$ of Dirac fermions carrying pseudo-real representations coupled to fundamental scalars, and impose P, C so as to guarantee $M_\phi=M_\phi^*$. Suppose in addition the theory enjoys a {flavor symmetry} in which at least one element of the flavor group is represented by an antisymmetric matrix, i.e. $\Psi_i\to U_{ij}\Psi_j$ with $U_{ij}=-U_{ji}$. Under this hypothesis one can replace $\eta {\cal T}$ by $U\eta {\cal T}$ everywhere in our argument and prove positivity, since this way ${\mathsf T}^2=-1$ as it was for real representations. However, this program is effectively equivalent to the previous one because the flavor group we are imposing always contains the $Z_4$ subgroup alluded to earlier.

\section{Applications}
\label{sec:appl}

Having established positivity of the fermionic determinant in a new class of theories with Yukawa couplings, we are now ready to discuss some concrete implications. As in \cite{Weingarten:1983uj,Vafa:1983tf,Vafa:1984xg}, by P invariance all topological angles $\theta$ vanish in the field basis in which $M_\phi$ is real. The full path integral is real and positive definite: 
\ba
\sum_\nu\int{\cal D}[A_E,\phi_E]~e^{-S_E^{\rm bosons}}{\rm det}[H]>0, 
\ea
with $A_E,\phi_E$ identifying the (Euclidean) bosonic variables and $e^{-S_E^{\rm bosons}}$ the exponential of the bosonic action (including gauge-fixing, if needed), which is real and positive definite. A sum over topological sectors is also included. It is important to stress that the above relation is a strict inequality: the path integral is positive definite, i.e. it does not vanish identically.~\footnote{Given \eqref{theorem}, to convince oneself of this it is sufficient to find a subset of bosonic configurations for which the equation $({\slashed{D}}+M_\phi)\chi=0$ admits no solutions. An example is the subset of ``small" bosonic fields. Indeed, suppose a solution $\chi$ of the equation $({\slashed{D}}+M_\phi)\chi=0$ existed. The zero-mode would then also solve $(-{\slashed{D}}+M_\phi)({\slashed{D}}+M_\phi)\chi=(-D^2+M_\phi^2-F)\chi=0$, where $F=\frac{1}{2}[{\slashed{D}},{\slashed{D}}]+\gamma^\mu\partial_\mu M_\phi$ is a function of the bosonic fields. Yet, by focusing on the contribution to the path integral restricted to bosonic configurations small enough such that $M_\phi^2-F>0$ everywhere, the latter equation cannot be satisfied because $-D^2\geq0$. Hence, the so-defined subset of ``small" bosonic configurations does not support fermionic zero-modes and (at least) their contribution to the path integral is strictly positive.} This is reassuring, otherwise one would be dealing with a pathological theory.

We will prove the conservation of parity in Section \ref{sec:Pconservation}, and the conservation of vector-like flavor symmetries in Section \ref{sec:Flavorconservation}. In Section \ref{sec:chiralLag} we will show that a chiral Lagrangian analysis is consistency with our claims. Finally, mass inequalities constraining the spectrum of our theories are derived in Section \ref{sec:massineq}. 

It is important that throughout our discussion the scalars $\phi$ are taken to be singlets under the global symmetry of interest, e.g. parity or vector-like symmetries. On the contrary, a sufficiently large and negative mass squared in the scalar potential would trigger spontaneous symmetry breaking compatibly with all our general requirements. Thus, in the presence of scalars with global charges our arguments do not apply and a quantitative model-dependent analysis is needed to establish whether symmetry breaking takes place. On the other hand, our results apply irrespective of the gauge quantum numbers of $\phi$.

Our proofs heavily rely on the path integral formalism. It is therefore important to emphasize that this technique should be employed with care when investigating spontaneous symmetry breaking. 

When the vacuum is unique the path integral is exactly delivering a time-ordered vacuum correlator, up to an irrelevant normalization. In the presence of multiple degenerate vacua $|\Omega^\alpha\rangle$ the path integral is a linear combination of matrix elements of the different vacua. Consider for instance the path integral of fields $\Phi$ subject to boundary conditions $\Phi(t_{i,f},{\vec x})=\Phi_{i,f}({\vec x})$ at the initial $t_i=-T/2$ and the final time $t_f=+T/2$, with the insertion of a local operator $O(x)$ evaluated at a time $t_i<x^0<t_f$. If, as usual, we momentarily continue the expression to Euclidean space and let $T\to\infty$ to project onto the vacuum states $|\Omega^\alpha\rangle$, one finds
\ba\label{pathphi}
\int_{\Phi_i}^{\Phi_f}{\cal D}\Phi~e^{iS}O(x)
&=&\langle\Phi_f|e^{-iHT/2}~O(x)~e^{-iHT/2}|\Phi_i\rangle\\\no
&\to&e^{-iE_0T}\sum_{\alpha,\beta}\langle\Phi_f|\Omega_{\rm out}^\alpha\rangle\langle\Omega_{\rm out}^\alpha|O(x)|\Omega_{\rm in}^\beta\rangle\langle\Omega_{\rm in}^\beta|\Phi_i\rangle.
\ea

By definition, spontaneous symmetry breaking singles out a specific vacuum out of the set $|\Omega^\alpha\rangle$. One would then be interested in computing the matrix elements of the $O$'s on that specific vacuum, say in order to compute the expectation value of an order parameter. As suggested by the theory of ferromagnets, we can achieve that by perturbing the theory with external sources $J$ (which we take to be constant in space-time for simplicity) coupled to some order parameters $X$, defined to be local operators transforming non-trivially under the broken symmetry and having non-vanishing expectation values on the $|\Omega^\alpha\rangle$'s:
\ba\label{pathphi11}
Z[J]\equiv\int_{\Phi_i}^{\Phi_f}{\cal D}\Phi~e^{iS-iJ\int X}.
\ea
With such a perturbation the energy degeneracy is lifted by corrections of order $\Delta E={\cal O}(JV_3\langle X\rangle)$.~\footnote{At leading non-trivial order in $J$, the correction to the Hamiltonian induced by a perturbation $\delta L(q,\dot q)=JL_1=J\int {\rm d}^3x\,X$ of the Lagrangian is controlled by $\delta H(q,p)=-JL_1+\cdots$. This is always true, even if $L_1$ depends on the variables $\dot q$. This justifies the statement $\Delta E={\cal O}(JV_3\langle X\rangle)$.} Hence, in the infinite volume limit $V_3T\to\infty$ a unique vacuum matrix element survives, the contribution of the others being exponentially suppressed by $e^{-|{\cal O}(JTV_3\langle X\rangle)|}$. When computing the vacuum matrix elements, the infinite volume limit must thus be performed {\emph{before}} switching off the source. Only in this way
\ba\label{CorrelatorsCSB}
\langle OO'\cdots\rangle=\lim_{J\to0}\lim_{V_4\to\infty}\frac{\int_{\Phi_i}^{\Phi_f}{\cal D}\Phi~OO'\cdots e^{iS-iJ\int X}}{\int_{\Phi_i}^{\Phi_f}{\cal D}\Phi~e^{iS-iJ\int X}}
\ea
correctly reproduces the vacuum correlators on the (true) vacuum that the source has singled-out. In particular, the vacuum expectation value of the order parameter is
\ba\label{standard<O>}
\langle X\rangle=\lim_{J\to0}\lim_{V_4\to\infty}\frac{i}{TV_3}\frac{\rm d}{{\rm d} J}\ln Z[J].
\ea 
Of course there would be no ambiguity in the order of the limits if the vacuum was unique. That $\lim_{TV_3\to\infty}$ must be taken before switching off the source is a consequence of the familiar fact that spontaneous symmetry breaking can only occur at infinite volume. 

After this introductory comment we are ready to present our results.

\subsection{Conservation of P}
\label{sec:Pconservation}

In this section we show that the proof of P-conservation (or equivalently CT-conservation) presented in \cite{Vafa:1984xg} for massive QCD-like theories goes through basically unchanged in our scenarios (where $M_\phi$ is a function of no definite sign, see \eqref{Mphi}) whenever the order parameter is constructed with purely bosonic fundamental fields and the $\phi$'s are P-even. 

Order parameters containing  fermionic fields are more difficult to treat. Still, we will see that with a mild modification of the argument of \cite{Vafa:1984xg} we can extend the original claim to arbitrary order parameters.~\footnote{Some criticism of the original Vafa-Witten argument can be found in \cite{Sharpe:1998xm,Azcoiti:1999rq,Cohen:2001hf,Ji:2001sa}. See also \cite{Einhorn:2002rm} for a rebuttal of some of them.} More precisely, we will demonstrate that P conservation is guaranteed if and only if the operator ${\slashed{D}}+M_\phi$ has no zero-modes. This implies that P is exactly conserved in QCD with $m>0$. Unfortunately, in our models with Yukawa couplings the existence of a gap in the spectrum of $H$, or equivalently P, cannot be rigorously proved unless some additional input is provided. To be able to make concrete statements we will thus invoke Gell-Mann's ``totalitarian principle": if P is spontaneously broken then any P-odd condensate should form unless prohibited by other (conserved) symmetries. This additional input is enough to argue that P is exactly conserved in all our models with P-even scalars. As we will see, the same principle was also implicitly assumed in \cite{Vafa:1983tf}.

In order to present a self-contained discussion we will first review the argument of \cite{Vafa:1984xg} for order parameters constructed with solely bosonic fields. Subsequently we will highlight the subtleties encountered when considering fermionic operators and present our approach.

\subsubsection*{Order parameters made of solely bosonic fields}

Suppose we consider a P-invariant theory of real vector fields $A_\mu$ and real P-even scalars $\phi$ coupled to Dirac fermions $\Psi,\overline\Psi$ in real representations of the gauge group. The theory is defined by a Lorentz-signature action $S=\int {\rm d}^4x~{\cal L}$. We follow Section \ref{sec:Flavorconservation} and perturb the theory adding a set of hermitian P-odd order parameters $X(x)$, of arbitrary engineering dimension, composed of only fundamental bosonic field variables multiplied by real constant sources $J$. The path integral in Minkowski space is formally given by 
\ba\label{PIappendix}
{\cal N}e^{-iTV_3\varepsilon(J)}=\sum_\nu\int {\cal D}[{\rm fields}]~e^{iS-iJ\int {\rm d}^4x\,X(x)+{\cal O}(J^2)}
\ea
with ${\cal N}$ a normalization. Importantly, in order to regularize the UV divergences in the path integral \eqref{PIappendix} we allowed counterterms of higher powers of $J$ (these are necessary unless $X$ is linear in the fundamental fields, which cannot be the case here). Because of this, \eqref{PIappendix} is not to be interpreted as a generating functional of arbitrary $X$ insertions.~\footnote{For the same reason in general there exists no notion of 1PI effective action. In this sense we disagree with the argument presented in \cite{Einhorn:2002rm}.} However this is of no concern to us because here we are only interested in the single insertion, i.e. linear order in $J$. In fact our focus is entirely on
\ba\label{dwdJ}
\langle X(0)\rangle=\lim_{J\to0}\frac{\partial \varepsilon}{\partial J}.
\ea
This can be unambiguously interpreted as the vacuum expectation value of the order parameter provided the operators $X$ form a closed set under renormalization. As reviewed in Section \ref{sec:Flavorconservation}, $\langle X(0)\rangle$ represents the average on the special vacuum selected by the perturbation. Importantly, by P invariance of the action and the measure the path integral is even in the source:
\ba\label{|J|}
\varepsilon(J)=\varepsilon(-J).
\ea
When P is spontaneously broken by the vacuum of $X$ the partition function depends on $|J|$ and is non-analytic in $J=0$ (implicitly we have already taken the infinite volume limit). The limit in \eqref{dwdJ} therefore depends on ${\rm sign}[J]$. We will come back to this crucial point below.

Our goal is to show that the quantity in \eqref{dwdJ} vanishes. To do this we rotate to Euclidean space $x=(t,\vec{x})\mapsto x_E=(-it_E,\vec{x})$. In doing so $iS\mapsto-S_E$, $iT\mapsto T_E$, and crucially $X(x)\mapsto iX_E(x_E)$, with $X_E(x_E)$ real. Indeed, by Lorentz invariance any P-violating combination of $A_\mu$ must contain odd powers of the Levi-Civita tensor, and acquire an imaginary $i$ when rotated to Euclidean space \cite{Vafa:1984xg}. The same conclusion obviously extends to theories with P-even scalars. Hence the Euclidean path integral, after having integrated out the fermions, reads
\ba\label{EuclideanPI}
{\cal N}_Ee^{-V_4\varepsilon(J)}
=\sum_\nu\int {\cal D}[A_E,\phi_E]~{\rm det}[H]~e^{-S^{\rm bosons}_E-iJ\int {\rm d}^4x_E\,X_E(x_E)+{\cal O}(J^2)},
\ea
with $V_4=T_EV_3$. Because the P-odd phase can only decrease the integral, positivity of the Euclidean path integral measure in the $|J|\to0$ limit guarantees that 
\ba\label{El>E0}
\varepsilon(J)\geq \varepsilon(0)+{\cal O}(J^2).
\ea
Crucially, ref. \cite{Vafa:1984xg} points our that such relation cannot possibly hold if spontaneous P violation is assumed to take place, as we will review shortly. Hence, the hypothesis of spontaneous P breaking cannot be correct. Note that our formulation differs slightly from the original one presented in \cite{Vafa:1984xg}, where it was stated that $\varepsilon(J)\geq \varepsilon(0)$. Following our logic, counterterms of order $J^2$ are generically present, and in principle can spoil the inequality.

A proof of the incompatibility of the P-violation hypothesis with \eqref{El>E0} can easily be established. To begin, let us assume that spontaneous P violation takes place. Hence $\varepsilon(J)=\varepsilon(0)+J\langle X\rangle+{\cal O}(J^2)$ has a term linear in $J$, see \eqref{dwdJ}. As reviewed at the beginning of Section \ref{sec:Flavorconservation}, the actual vacuum $|\Omega_J\rangle$ is $J$-dependent but, by hypothesis, $\langle X\rangle\equiv\lim_{J\to0}\langle\Omega_J|X|\Omega_J\rangle$ is ambiguous because there exist multiple vacua of the unperturbed theory, i.e. the limit $\lim_{J\to0}|\Omega_J\rangle=|\Omega_0\rangle$ is not unique. The $|\Omega_0\rangle$'s must occur in P-conjugated pairs, and it is always possible to define two linear combinations $|\Omega_0^\pm\rangle$ of the $|\Omega_0\rangle$'s that are mutually orthogonal and such that $\langle\Omega_0^\pm| X(0)|\Omega_0^\pm\rangle=\pm\epsilon_X$ with $\epsilon_X$ real. Hence, for any sign of $J$ we can always find a $J=0$ vacuum for which $\varepsilon(J)=\varepsilon(0)+J\langle X\rangle+{\cal O}(J^2)=\varepsilon(0)-|J\epsilon_X|+{\cal O}(J^2)\leq \varepsilon(0)+{\cal O}(J^2)$. But such conclusion would lead to a contradiction of \eqref{El>E0} at small $|J|$ unless $\epsilon_X=0$. Therefore, the hypothesis of spontaneous P-violation must be false.~\footnote{Superficially there would seem to be no contradiction of \eqref{El>E0} if $\langle X(0)\rangle=0$. But if one can show that the above argument holds for {\emph{any}} hermitian order parameter $X(x)$, then one must conclude that either P is not spontaneously broken or any order parameter must have vanishing vacuum expectation value, which are two physically equivalent statements.} Note that this argument remains valid irrespective of gauge-invariance and therefore independently of the gauge quantum numbers of $\phi$. On the other hand, positivity cannot certainly be enough to ensure symmetry conservation if the fundamental scalars are P-odd, as emphasized at the beginning of Section \ref{sec:appl}.

\subsubsection*{Order parameters with fermionic fields}

From the previous discussion follows that whenever eq. \eqref{El>E0} holds, then P cannot be spontaneously broken. For order parameters made up of only bosonic fields \eqref{El>E0} was obvious. The same conclusion also extends to a class of order parameters for which the action can be expressed as in \eqref{PIappendix} via field re-definitions. In the case of $X=\overline{\Psi}i\gamma_5M_\phi\Psi$, for example, after an anomalous chiral transformation one ends up with a Lagrangian that formally looks like the original one except for the presence of ${\cal O}(J^2)$ corrections in the fermonic part and the addition of a topological term with ${\cal O}(J)$ source. The former describe subleading interactions in the Hamiltonian that can only affect the vacuum energy density $\varepsilon(J)$ at ${\cal O}(J^2)$ and hence do not impact our earlier argument. We are thus allowed to re-run it with $X=({g^2}/{32\pi^2})F^A\widetilde F^A$ and conclude that (in Minkowski signature)
\ba\label{opP}
\langle\overline{\Psi}i\gamma_5M_\phi\Psi\rangle=-2T_{\cal R}\frac{g^2}{32\pi^2}\langle F^A\widetilde F^A\rangle.
\ea
According to our earlier results, they must both vanish.

Unfortunately, things are not that simple when $X$ is a generic operator constructed with fermionic fields because in that case one generically encounters IR singularities associated to potential fermionic zero-modes, as we will see shortly. Without additional assumptions we cannot claim we have a rigorous proof of P conservation. As stated at the beginning of this section, in this work the additional assumption is a reasonable physical principle: unless otherwise prohibited by some symmetry, all vacuum expectations values must form. According to this hypothesis, because \eqref{opP} as well as all other P-odd operators constructed with only bosonic fields are singlets of all symmetries except P, there would be no reason for them to have a vanishing vacuum expectation value besides P-conservation. This physical consideration, and the phenomenological analysis in Section \ref{sec:chiralLag}, will be interpreted as evidence that P remains conserved in all our models with P-even scalars.

It is very instructive to illustrate the subtleties that emerge when studying generic order parameters with fermionic fields. Consider the P-violating operator $X=\overline{\Psi}i\gamma_5\Psi$. Integrating over the fermions in the (Euclidean) path integral we formally obtain 
\ba\label{PIfermP}
{\cal N}_Ee^{-V_4\varepsilon(J)}
&=&\sum_\nu\int {\cal D}[{\rm fields}_E]~e^{-S_E-J\int \overline{\Psi}i\gamma_5\Psi}\\\no
&=&\sum_\nu\int {\cal D}[A_E,\phi_E]~{\rm det}[H+iJ]e^{-S_E^{\rm bosons}}.
\ea
We have to be very careful in evaluating the derivative with respect to $J$ because as we will see there are subtleties associated with the zero-modes of $H$. The only P-conserving IR regulator that we can think of is the volume itself. Hence we will work at finite $V_4$. In that regime the spectrum of $H$ is discrete and the modulus of the non-zero eigenvalues is bounded from below by some $V_4$-dependent IR cutoff. The vacuum expectation value of the (possible) order parameter is obtained by differentiating with respect to $J$ the path integral, integrating over the bosonic fields, dividing by the four-volume, sending $V_4\to\infty$, and finally taking the $J\to0$ limit. Observe that
\ba\label{psi5psiVEV1}
\frac{\partial}{\partial J}\ln{\rm det}[H+iJ]=\sum_{\lambda_k}\frac{i}{\lambda_k+iJ}=\sum_{\lambda_k}\left[\frac{J}{\lambda_k^2+J^2}+i\frac{\lambda_k}{\lambda_k^2+J^2}\right]
\ea
where $\lambda_k$ are the eigenvalues of $H$ at finite volume. Defining now the bosonic average of the density of eigenvalues per unit volume (i.e. the spectral density of $H$),
\ba\label{rho}
\rho(\lambda)\equiv\lim_{J\to0}\lim_{V_4\to\infty}\frac{\sum_\nu\int{\cal D}[A_E,\phi_E]~e^{-S_E^{\rm bosons}}{\rm det}[H+iJ]~\frac{1}{V_4}\sum_{k}\delta(\lambda-\lambda_k)}{\sum_\nu\int{\cal D}[A_E,\phi_E]~e^{-S_E^{\rm bosons}}{\rm det}[H+iJ]},
\ea
the vacuum expectation value we are interested in can be written as 
\ba\label{psi5psiVEV2}
-\langle\overline{\Psi}i\gamma_5\Psi\rangle&=&\lim_{J\to0}\int {\rm d}\lambda\,\rho(\lambda)\left[\frac{J}{\lambda^2+J^2}+i\frac{\lambda}{\lambda^2+J^2}\right]\\\no
&=&-{\rm sign}[J]\,\pi\rho(0)+i\int {\rm d}\lambda\,\rho(\lambda)~{\cal P}\left(\frac{1}{\lambda}\right),
\ea
where we used $\lim_{\epsilon\to0}{1}/({\lambda+i\epsilon})=\lim_{\epsilon\to0}(-i\epsilon+\lambda)/({\lambda^2+\epsilon^2})=-{\rm sign}[\epsilon]\,\pi\delta(\lambda)+{\cal P}(1/\lambda)$, with ${\cal P}$ the Cauchy principal value.

Let us pause for a moment to inspect \eqref{psi5psiVEV2}. To begin, note that $\rho(\lambda)$ is real, irrespective of whether P is spontaneously broken or not. In fact, the explicit expression of $\rho^*(\lambda)$ is precisely the same as that of $\rho(\lambda)$ provided $J\to-J$; but, as we saw around \eqref{|J|}, the path integral is always an even function of $J$, and so $\rho^*(\lambda)=\rho(\lambda)$. Furthermore, the integral of the principal value in \eqref{psi5psiVEV2} must vanish. There are at least two intuitive reasons why this should be the case. First, as we have emphasized earlier, any order parameter should have a vacuum expectation proportional to ${\rm sign}[J]$, as the first term in \eqref{psi5psiVEV2}. And yet the integral of the principal value seems to carry no information about the sign of $J$. Furthermore, the vacuum expectation value $\langle\overline{\Psi}i\gamma_5\Psi\rangle$ is real in Minkowski space, but the integral would introduce a purely imaginary contribution. The only way to reconcile our expression \eqref{psi5psiVEV2} with these expectations is if the integral of the principal value in \eqref{psi5psiVEV2} vanishes. Indeed, the reason why $\int {\rm d}\lambda\,\rho(\lambda)~{\cal P}({1}/{\lambda})=0$ is that the (averaged!) density of eigenstates is an even function. To show this observe that determining $\rho(-\lambda)$ is equivalent to calculating $\rho(\lambda)$ with $H\to-H$. Now, because all eigenvalues of $H$ have even degeneracy we know that ${\rm det}[-H+iJ]={\rm det}[H-iJ]$, so changing the sign of $H$ is equivalent to changing the sign of $J$. Hence we are back to a situation similar to the one just discussed: by P invariance we can replace $J\to-J$ without affecting the result. In formulas, what we just argued is that $\rho(-\lambda)=\rho(\lambda)$ and
\ba\label{psi5psiVEV3}
\langle\overline{\Psi}i\gamma_5\Psi\rangle={\rm sign}[J]\,\pi\rho(0).
\ea

At this point we are in a position to make a very interesting statement: {\emph{P remains unbroken if and only if the operator ${\slashed{D}}+M_\phi$ has no zero-modes in the infinite volume limit}}. 

That P conservation implies the absence of zero-modes follows from \eqref{psi5psiVEV3} and the considerations of Section \ref{sec:appl}. Consistently with the latter, in a P-conserving theory any correlator function can be calculated without adding the source coupled to $X$. In particular, the (vanishing) vacuum expectation value of the P-odd biliear is given by eq.\eqref{psi5psiVEV2} with $J=0$. In other words, when P is unbroken the limits $J\to0$ and $V_4\to\infty$ are interchangeable. Taking the limit $J\to0$ in \eqref{rho} first, the quantity $\rho(\lambda)$ is seen to be the average of positive quantities weighted with a strictly positive measure, and in order to vanish the integrand must be zero as well. Therefore, when P is conserved, $\rho(0)=0$ implies that $H$ cannot have zero-modes. 

The opposite is also true. Suppose that $H$ has no zero-modes. In particular, suppose the non-trivial spectrum is bounded from below $|\lambda_k|\geq\lambda_{\rm min}$ for any $V_4$. In this situation one can expand ${\rm det}[H+iJ]$ perturbatively in $J/\lambda_{\rm min}$ and recover, up to ${\cal O}(J^2)$, a path integral completely analogous to the one analyzed in \eqref{EuclideanPI}, except for the fact that the order parameter has turned into a functional of bosonic fields. Explicitly, in the absence of zero-modes we have 
\ba\label{detHJ}
{\rm det}[H+iJ]
&=&\prod_{\lambda_k\neq0}(\lambda_k+iJ)\\\no
&=&e^{\sum_{\lambda_k\neq0}\ln(\lambda_k+iJ)}\\\no
&=&e^{\sum_{\lambda_k\neq0}\frac{1}{2}\ln(\lambda_k^2+J^2)+i\left[\sum_{\lambda_k\neq0}\arctan\frac{J}{\lambda_k}+\pi\,{\rm sign}(J)\sum_{\lambda_k<0}\right]}\\\no
&=&e^{\sum_{\lambda_k\neq0}\frac{1}{2}\ln(\lambda_k^2+J^2)+i\sum_{\lambda_k\neq0}\arctan\frac{J}{\lambda_k}},
\ea
where in the last step we used the fact that the multiplicity of the eigenstates is even. The presence of a mass gap at any $V_4$ ensures that~\footnote{At finite volume $H$ is an $N\times N$ matrix with $N=V_4/\ell^4$, see footnote \ref{foot}. The proportionality to $V_4$ in the reminder indicates that the free energy is an extensive quantity.}
\ba\label{lndetJ}
\ln{\rm det}[H+iJ]
&=&\ln {\rm det}[H]+\sum_{\lambda_k\neq0}\left\{\frac{1}{2}\ln(1+J^2/\lambda_k^2)+i\arctan\frac{J}{\lambda_k}\right\}\\\no
&=&\ln {\rm det}[H]+i\sum_{\lambda_k\neq0}\frac{J}{\lambda_k}+\frac{V_4}{\ell^4}{\cal O}(J^2/\lambda_{\rm min}^2)
\ea
which confirms we are indeed in the situation captured by eq. \eqref{EuclideanPI} with
\ba
\int {\rm d}^4x_E\,X_E(x_E)=-\sum_{\lambda_k\neq0}\frac{1}{\lambda_k}=-{\rm Tr}\frac{1}{H}.
\ea
(This is a complicated functional of the bosonic fields whereas in QCD it is just the topological charge, see Appendix \ref{sec:masslessQCD}.) This is enough to conclude that $\langle\overline{\Psi}i\gamma_5\Psi\rangle=0$. Incidentally, eq. \eqref{lndetJ} also shows that when fermionic order parameters are considered $\varepsilon(J)\geq\varepsilon(0)$ may not be correct, as it is evident that $|{\rm det}[H+iJ]|\geq{\rm det}[H]$. Nevertheless, \eqref{El>E0} continues to hold up to $O(J^2)$.~\footnote{In the literature there are claims that because $\varepsilon(J)\geq\varepsilon(0)$ does not hold in general, then the arguments of \cite{Vafa:1984xg} do not apply to order parameters with fermions. We have provided ample evidence that this is incorrect. The proof of P conservation does not require $\varepsilon(J)\geq\varepsilon(0)$, but rather the weaker condition \eqref{El>E0}.} An analogous perturbative expansion in $J^2$ can be performed in the calculation of the vacuum expectation value of any P-odd $X$. We conclude that P-conservation can be established for any order parameter as we did for bosonic $X$'s as long as the spectrum of $H$ has a mass gap in the infinite volume limit, as promised.

Our earlier claim of exact P conservation, see around \eqref{opP}, motivated by a few rigorous considerations combined with the ``totalitarian principle", is thus physically equivalent to the statement that $H$ has no zero-modes. Should we have expected that? It is certainly true that zero-modes are absent if $m\neq0$ and the spectrum of eigenvalues is analytic in $y$ around $y=0$. However a priori we see no reason for this to be the case in general and are not able to provide a direct proof. What we can more humbly verify is that the absence of zero-modes for $H$ is not in conflict with the index theorem \eqref{indexTH}, as the latter relates the topological index to the number of zero-modes of the operator ${\slashed{D}}$, not $H$. Nevertheless, we should further clarify our claim. From the anti-hermiticity of ${\slashed{D}}$ follows that a hypothetical zero-modes $\chi_0$ must satisfy $\int \chi_0^\dagger M_\phi\chi_0=0$. Hence, in principle a non-trivial $\chi_0$ might exist even if $m\neq0$ provided $y\neq0$. An obvious solution can be found if the scalar is chosen to be exactly $\phi=-m/y$, so that $M_\phi=0$. However, that configuration does not affect the path integral because it has vanishing measure in field space. We can imagine removing that --- together with all other isolated configurations leading to $H\chi_0=0$, if present --- without affecting physics. It is with this caveat that, we claim, $H$ admits no zero-modes.

Before concluding, it is useful to compare our models to QCD (see Appendix \ref{sec:masslessQCD} for more details). In massive QCD the spectrum of the eigenvalues of ${\slashed{D}}+m$ is obviously gapped and as we argued above this ensures that given an arbitrary order parameter the bosonic integrand contains no ${\cal O}(J)$ terms up to a phase precisely as in \eqref{EuclideanPI}. Therefore one can safely conclude that parity is exactly conserved, provided of course the path integral has a positive measure ($m>0$). This is precisely the framework assumed by ref. \cite{Vafa:1984xg}. From a symmetry standpoint our scenarios are completely analogous. Our claims that $H$ has no zero-modes and P is conserved are also compatible with what happens in massive QCD.

\subsection{Conservation of vector-like flavor symmetries}
\label{sec:Flavorconservation}

In this section we will show that the vector-like flavor symmetries, by which we mean the subgroup of the $SU(2N_f)$ fermion flavor rotations that are preserved by the fermion mass and Yukawa couplings, are not spontaneously broken. We will first show that the vacuum expectation value of order parameters for all vector-like symmetries, including discrete ones, vanish. Subsequently we will discuss an obstruction we encountered in adapting to our models an argument proposed in \cite{Vafa:1983tf} that excludes the existence of massless Nambu-Goldstone bosons associated to the breaking of vector-like continuous symmetries.

\subsubsection*{Vanishing expectation value of the order parameters}  
\label{sec:VEVargument}

We would like now to present an argument that constrains the breaking of fermionic global symmetries. Following the source formalism reviewed at the beginning of Section \ref{sec:Flavorconservation}, we aim to show that the order parameters for vector-like symmetries must have vanishing vacuum expectation value whenever the operator ${\slashed{D}}+M_\phi$ has a mass gap. To make our discussion concrete we go back to the model \eqref{pseudoL} with gauge-singlet scalars. Yet, considerations analogous to the ones presented below can be employed to argue that vector-like symmetries (continuous and discrete) are exactly conserved in all models discussed in Section \ref{sec:gene}, including models with more flavors or with gauge-charged, flavor-neutral scalars.

The model \eqref{pseudoL} has a $U(1)_B$ flavor symmetry for which a possible order parameter is
\ba
{\mathbb M}_1(x)=\overline{\Psi}(x)(\eta {\cal T})\overline{\Psi}^t(x)
\ea
or its hermitian (in Minkowski) conjugate $\overline{{\mathbb M}}_1=\Psi^t(\eta {\cal T})^\dagger\Psi$. According to \eqref{pathphi11}, we are interested in the perturbed Euclidean path integral (for brevity the sum over topological sectors is left understood in this section)
\ba\label{ZJJbar}
Z_E[J,\overline{J}]
&=&\int{\cal D}[{\rm fields}_E]~e^{-S_E^{\rm bosons}-\int {\rm d}^4x\,\overline{\Psi}\gamma_5H\Psi-J\int {\rm d}^4x\,{\mathbb M}_1-\overline{J}\int {\rm d}^4x\,\overline{{\mathbb M}}_1}\\\no
&=&C\int{\cal D}[A_E,\phi_E]~e^{-S_E^{\rm bosons}}\int{\rm d}[\bar b,a]~e^{-\sum_n\lambda_n\bar b_na_n-J\sum_{m,n}c_{mn}\bar b_m\bar b_n-\overline{J}\sum_{m,n} c^*_{mn}a_ma_n}.
\ea
In the second line we have expanded the fermionic fields as $\Psi=\sum_na_n\chi_n$, $\overline\Psi=\sum_n\bar b_n\chi_n^\dagger\gamma_5$, with $\chi_n$ the orthonormalized eigenvectors of $H$ with eigenvalue $\lambda_n$, and performed the change of basis $\Psi\to a_n$, $\overline\Psi\to\bar b_n$, with $C$ a field-independent Jacobian. The order parameter becomes
\ba
&&\int {\rm d}^4x~{\mathbb M}_1=\sum_{m,n}\bar b_m\bar b_n\int {\rm d}^4x~\chi_m^\dagger(\eta {\cal T})\chi_n^*=\sum_{m,n}\bar b_m\bar b_n c_{mn}\\\no
&&\int {\rm d}^4x~\overline{\mathbb M}_1=\sum_{m,n}a_ma_n\int {\rm d}^4x~\chi_m^t(\eta {\cal T})^\dagger\chi_n=\sum_{m,n}a_ma_n c^*_{mn}
\ea
where $c_{mn}$ is a unitary anti-symmetric matrix defined in \eqref{Nf=1}, though here $n$ is extended to label all eigenvalues. We next explicitly compute the Grassmanian integral expanding the source-dependent term. The typical contribution reads:
\ba\label{ZJJbar1}
Z_E[J,\overline{J}]
=\int{\cal D}[A_E,\phi_E]~e^{-S_E^{\rm bosons}}\prod_k\lambda_k\left[1-2J\overline{J}\sum_{m,n}\frac{c_{mn} c^*_{mn}}{\lambda_m\lambda_n}+\cdots\right].
\ea
Crucially, invariance of the action under the flavor symmetry implies that in order to obtain a non-vanishing result the same number of $J$ and of $\overline{J}$ is needed. The source-dependent term formally represents a series in ${J\overline{J}c c^*}/{\lambda^2}$, with the higher order terms indicated by the dots.

Now, eq. \eqref{ZJJbar1} superficially seems to indicate that the vacuum expectation value of the order parameter vanishes as the sources are switched off. But this naive conclusion is invalidated when some of the eigenvalues approach zero in the infinite volume limit. In that case the analytic expansion in ${J\overline{J}c c^*}/{\lambda^2}$ simply makes no sense. For example, when $\rho(0)$ is a non-trivial constant the lightest eigenvalues scale as $\lambda\sim1/(\rho(0)V_4)$ and the naive series in the small parameter $J$ turns into an expansion in $JV_4\rho(0)$, which in fact diverges.

Fortunately, we have argued earlier that our models avoid this problem because $H$, or equivalently ${\slashed{D}}+M_\phi$, has no zero-modes. Hence the absolute values of the eigenvalues $\lambda_n$ are bounded from below by a field-independent quantity, i.e.
\ba\label{hypoth}
|\lambda_n|\geq\lambda_{\rm min}\neq0,
\ea
for any $V_4$. Then, using positivity of the measure the expression, \eqref{ZJJbar1} is shown to imply 
\ba
\left|\frac{\partial Z_E[J,\overline{J}]}{\partial J}\right|&\leq&\lim_{J,\overline{J}\to0}\int{\cal D}[A_E,\phi_E]~e^{-S_E^{\rm bosons}}\left|\prod_k\lambda_k~\overline{J}\sum_{m,n}2\frac{c_{mn} c^*_{mn}}{\lambda_m\lambda_n}\right|+\cdots\\\no
&\leq&\lim_{J,\overline{J}\to0}\int{\cal D}[A_E,\phi_E]~e^{-S_E^{\rm bosons}}\left|\prod_k\lambda_k\right||\overline{J}|\sum_{m,n}2\frac{c_{mn} c^*_{mn}}{\lambda_{\rm min}^2}+\cdots\\\no
&=&Z_E[0,0]\,2\frac{|\bar J|}{\lambda_{\rm min}^2}\frac{V_4}{\ell^4}+\cdots
\ea
where the first inequality follows from taking the absolute value inside the integral, the second from taking it inside the sum and our hypothesis \eqref{hypoth}, and the last equality is a consequence of positivity and the relation $\sum_{m,n}{c_{mn} c^*_{mn}}=\sum_n=V_4/\ell^4$.~\footnote{\label{foot}On a lattice of cutoff length $\ell$ we can interpret $\chi_n(\ell\hat x)$ as an $N\times N$ matrix satisfying the completeness relations $\sum_{\hat x}(\chi^\dagger\chi)_{mn}=\delta_{mn}/\ell^4$ and $(\chi\chi^\dagger)_{\hat x\hat y}=\delta_{\hat x\hat y}/\ell^4$. Hence $V_4=\sum_{\hat x}\ell^4=N\ell^4$.} Analogously one can show that higher-order terms, indicated by the dots, scale with additional powers of $(J\bar J V_4/\lambda_{\rm min}^2\ell^4)$. Hence we get
\ba\label{VJres}
\frac{1}{V_4}\left|\frac{\partial Z_E[J,\overline{J}]}{\partial J}\right|\leq Z_E[0,0]\,\frac{2|\bar J|}{\lambda_{\rm min}^2\ell^4}~F_1\left(\frac{|J\bar J|V_4}{\lambda_{\rm min}^2\ell^4}\right)
\ea
as well as $|Z_E[J,\overline{J}]|\leq Z_E[0,0]F_0(|J\bar J|V_4/\lambda_{\rm min}^2\ell^4)$, for some unknown functions satisfying $F_0(0)=F_1(0)=1$. 

What can \eqref{VJres} say about the vacuum expectation value of the order parameter? Absence of spontaneous symmetry breaking would imply that the left hand side of \eqref{VJres} vanished if we first take the infinite volume limit and then send the source to zero, as in \eqref{standard<O>}. But we do not seem to be able to conclude anything concrete directly from \eqref{VJres} unless more information about the functions $F_{0,1}$ is provided. This is not completely correct however. Rather than taking $V_4\to\infty$ and then $J\to0$ separately, we propose to adopt a procedure alternative to \eqref{standard<O>} to compute the vacuum expectation value, which also takes into account the non-commutativity of the two limits: we assume the source scales with the volume as 
\ba
J\equiv J_0/V_4^\gamma
\ea
for $0<\gamma<1$ and take a unique limit $V_4\to\infty$. In this way, when $V_4$ is sent to infinity we are also automatically switching off the effect of the source, but we do so at a slower rate, as we should. The vacuum expectation value one obtains in this way coincides with the one derived via the more familiar expression \eqref{standard<O>}:
\ba\label{<O>conv}
\langle X\rangle=-\lim_{V_4\to\infty}\frac{1}{V_4}\left.\frac{\partial}{\partial J}\ln Z_E\right|_{J=J_0/V_4^\gamma}
\ea 
because for {\emph{any}} $0<\gamma<1$ the quantity ${\cal O}(JV_4\langle X\rangle)$ diverges and allows us to project out the would-be degenerate vacua except for a single one. So, the arguments reviewed at the beginning of Section \ref{sec:Flavorconservation} go through unchanged.~\footnote{As a concrete well-known example one may consider the infinite range Ising model with Hamiltonian $H=-\frac{g}{2N}\sum_{i,j=1}^N\sigma_i\sigma_j-J\sum_{i=1}^N\sigma_i$. The partition function can be calculated exactly in the infinite ``volume" limit $\overline{N}=N/(kT)\to\infty$ and reads $Z[J]=2\exp[\overline{N}g/2]\cosh[\overline{N}J]$. The correct average $\langle\sum\sigma_i\rangle/\overline{N}$ can be derived via the standard formula $\lim_{J\to0}\lim_{\overline{N}\to\infty}\frac{1}{\overline{N}}\partial\ln Z/\partial J$ or equivalently defining $J=J_0/{\overline{N}}^\gamma$ (with $0<\gamma<1$) and taking $\lim_{\overline{N}\to\infty}\frac{1}{\overline{N}}\partial\ln Z/\partial J$.} In the specific case at hand, taking $1/2<\gamma<1$ and the infinite volume limit as prescribed by eq. \eqref{<O>conv}, we see that $Z_E[J,\overline{J}]\to Z_E[0,0]$ and \eqref{VJres} vanishes, and hence $\langle{\mathbb M}_1\rangle=0$. It follows that ${\mathbb M}_1$ cannot be an order parameter for $U(1)_B$ breaking.

Our conclusion can be generalized to order parameters of arbitrary dimensionality provided $S+J\int X$ can be equivalently written as an action quadratic in the fermion fields and involving (auxiliary) bosons compatible with positivity of the measure (see for example the cases analyzed in Section \ref{sec:gene}). 

The analysis of order parameters with also bosonic fields is more subtle, though, because gauge fields have perturbative zero-modes (and in our case scalars can have tachyonic masses) and it is thus not obvious that the formal series in $J$ has a well-defined $J\to0$ limit. This subtlety should not invalidate our conclusions nor those of \cite{Vafa:1983tf} (and \cite{Weinberg:1996kr}), however. Intuitively, the perturbative zero-modes of the bosonic fields are lifted non-perturbatively in asymptotically-free gauge theories. Hence, our argument may be re-run sending the source to zero {\emph{after}} the integration over the bosonic fields is performed. Another reason why the above caveat should not represent a serious obstruction, physically, is that the possibility that order parameters formed solely by fermionic fields do not get a vacuum expectation value, consistently with our findings, while those formed by fermionic {\emph{and bosonic}} fields do seems very implausible. Hence, we are confident that the arguments presented here suffice to prove the absence of spontaneous symmetry breaking of global fermionic symmetries.

\subsubsection*{Absence of Nambu-Goldstone excitations?}  
\label{sec:NGBargument}

Vafa and Witten suggested in ref. \cite{Vafa:1983tf} a method to establish the absence of spontaneous breaking of {\emph{continuous}} fermionic symmetries. Interestingly, that approach is completely impervious to the ambiguity mentioned in the introduction of Section \ref{sec:appl}. If symmetry breaking of a continuous global symmetry takes place then Nambu-Goldstone bosons must appear and must couple to both the order parameter and the symmetry current. Denoting with $O(x)$ one of these operators, then ($x^0>y^0$ for simplicity)
\ba\label{pathphi1}
\int_{\Phi_i}^{\Phi_f}{\cal D}\Phi~e^{iS}O(x)O^\dagger(y)
\to e^{-iE_0T}\sum_{\alpha,\beta}\langle\Phi_f|\Omega_{\rm out}^\alpha\rangle\langle\Omega_{\rm out}^\alpha|O(x)O^\dagger(y)|\Omega_{\rm in}^\beta\rangle\langle\Omega_{\rm in}^\beta|\Phi_i\rangle
\ea
necessarily manifests a power-law dependence on $x-y$ due to the propagation of a Nambu-Goldstone mode, independently of whether an order parameter $X$ for that symmetry is added to the action. Thus, if one can show that the 2-point functions of order parameters and currents do not feature such a power-law behavior, then spontaneous symmetry breaking cannot take place. Crucially, this statement applies to all matrix elements $\langle\Omega^\alpha|O(x)O^\dagger(y)|\Omega^\beta\rangle$ (note that $\alpha\neq\beta$ in general) in the sum of \eqref{pathphi1} and hence holds for the path integral itself even though such a quantity does not represent {\emph{the}} vacuum correlator in a theory with spontaneous symmetry breaking. In other words, the criteria of \cite{Vafa:1983tf} for the absence of symmetry breaking can be applied directly to the path integral \eqref{pathphi1} and does not require the introduction of sources. However, because it relies on the existence of Nambu-Goldstone bosons, it can only constrain the breaking of continuous global symmetries.

Let us look at the argument in more detail. What the authors of \cite{Vafa:1983tf} actually proved is that in QCD with $m>0$ the propagator of (appropriately smeared) fermionic fields $\Psi$, $\overline\Psi$ localized at separated Euclidean points $x,y$ are bounded by a field-independent quantity $\propto e^{-m|x-y|}$. Positivity of the measure then guarantees that the same exponential behavior is found in the 2-point function of the order parameters and the 2-point function of the flavor currents as long as these are made {\emph{solely of fermionic fields}}. The analysis of operators with also bosonic fields suffers from the very same subtlety mentioned in the paragraph below eq. \eqref{<O>conv}. Physically, this subtlety should not invalidate the conclusion of \cite{Vafa:1983tf} for the same reason it does not invalidate ours, however. Hence, the exponential fall-off of the 2-point function of operators $O$ made of fermion fields only should be enough to argue in favor of the absence of Nambu-Goldstone modes and of spontaneous symmetry breaking of global continuous symmetries.

Unfortunately, one cannot straightforwardly extend the logic of \cite{Vafa:1983tf} to our theories. While positivity was proved earlier, the presence of Yukawa interactions implies the effective fermionic mass is not strictly bounded. Even taking into account that $H$, and hence ${\slashed{D}}+M_\phi$, has no zero-modes, in order to apply the results of \cite{Vafa:1983tf} to our theories we would need to show that the real part of the eigenvalues of ${\slashed{D}}+M_\phi$ are bounded below by a real field-independent quantity, which we were not able to do.

To overcome this hurdle we will attempt to slightly deform our theory. We add a P-violating mass term
\ba\label{defo}
\delta S^{\rm fermions}_E=\int {\rm d}^4x~im_5\overline\Psi\gamma_5\Psi, 
\ea
with $m_5$ a real matrix proportional to the identity in flavor space. With that perturbation the operator $H=\gamma_5({\slashed{D}}+M_\phi)$ is replaced by $H\to H+im_5$ and its eigenvalues are $\lambda_n+im_5$, with $\lambda_n$ real. According to \cite{Vafa:1983tf}, in order to demonstrate the exponential fall-off of the (smeared) fermion propagator it is sufficient to show that the matrix element $\langle \beta|({{\slashed{D}}+M_\phi+im_5\gamma_5})^{-1}|\alpha\rangle$, with $|\alpha\rangle$ and $|\beta\rangle$ two smeared position states of support separated by a distance $R$, is bounded by a field-independent exponential suppression proportional to $R$. In our case $({\slashed{D}}+M_\phi+im_5\gamma_5)^{-1}=(H+im_5)^{-1}\gamma_5$, and 
\ba\no
\langle \beta|\frac{1}{{\slashed{D}}+M_\phi+im_5\gamma_5}|\alpha\rangle
&=&\frac{1}{i\,{\rm sign}(m_5)}\int_0^\infty d\tau\,\langle \beta|e^{-\tau(|m_5|-i\,{\rm sign}(m_5)H)}|\alpha\rangle\gamma_5\\\no
&=&\frac{1}{i\,{\rm sign}(m_5)}\int_0^\infty d\tau\,e^{-|m_5|\tau}\langle \beta|e^{-i\tau(-{\rm sign}(m_5)H)}|\alpha\rangle\gamma_5.
\ea
As in \cite{Vafa:1983tf}, the correlator $\langle \beta|e^{-i\tau(-{\rm sign}(m_5)H)}|\alpha\rangle\propto\theta(\tau-R)$ because $-{\rm sign}(m_5)H$ can be viewed as a causal Hamiltonian in a fictitious $4+1$ dimensional space-time, with $\tau$ being identified with ``time". Mathematically, that can be understood noting that 
\ba\label{HypPDE}
\frac{\partial}{\partial\tau}\psi=i\,{\rm sign}(m_5)H\psi
\ea
is a hyperbolic partial differential equation. As well-known, hyperbolicity is established by inspecting the coefficient of the highest derivative term on the right-hand side of eq. \eqref{HypPDE}, which in our case is an effective Gamma matrix $\Gamma^\mu\equiv i\,{\rm sign}(m_5)\gamma_5\gamma^\mu$. Hyperbolicity is not affected by the non-derivative terms, and hence by the vector and scalar couplings. As a result, $\langle \beta|e^{-i\tau(-{\rm sign}(m_5)H)}|\alpha\rangle$ has support only for $\tau\geq R$ and one obtains the desired exponential suppression
\ba\label{VW1proofm_5}
\left\Vert\langle \beta|\frac{1}{{\slashed{D}}+M_\phi+im_5\gamma_5}|\alpha\rangle\right\Vert\leq \frac{e^{-|m_5|R}}{|m_5|}\sqrt{\langle \alpha|\alpha\rangle}\sqrt{\langle \beta|\beta\rangle},
\ea 
with $\left\Vert\right.$ denoting the matrix norm.

Armed with \eqref{VW1proofm_5}, the argument of \cite{Vafa:1983tf} appear to go through almost unchanged. Given a smeared operator $O$ made of a number $k$ of (smeared) fermionic pairs, in particular an order parameter or a symmetry current, the path integral is bounded by:
\ba\label{ratioPI}
&&\left\Vert\frac{\int{\cal D}({\rm fields})_E~OO'\cdots e^{-S_E}}{\int{\cal D}({\rm fields})_E~e^{-S_E}}\right\Vert\\\no
&\leq&\left(\frac{e^{-|m_5|R}}{|m_5|}\sqrt{\langle \alpha|\alpha\rangle}\sqrt{\langle \beta|\beta\rangle}\right)^k\frac{\int{\cal D}({\rm bosons})_E~\left|{\rm det}[H+im_5]\right| e^{-S_E}}{\left|\int{\cal D}({\rm bosons})_E~{\rm det}[H+im_5]e^{-S_E}\right|}.
\ea
The problems with this logic is that when $m_5\neq0$ the path integral measure is complex and positivity is lost. Yet, this is only true because of corrections suppressed by powers of $m_5^2$. Thus, naively, by taking $|m_5|$ sufficiently small but non-zero one would hope to conclude that the 2-point functions of order parameters and currents constructed {{solely of fermionic fields}} are bounded by powers of \eqref{VW1proofm_5} modulo small corrections.~\footnote{This erroneous view was presented in our original preprint.}

Unfortunately, those apparently small corrections are enhanced by the volume.~\footnote{This issue is analogous to the one encountered in simple approaches to the sign problem in lattice gauge theories.} As eq. \eqref{lndetJ} makes it clear (with $J=m_5$) one has $|{\rm det}[H+im_5]|={\rm det}[H]\,\exp\left[c m_5^2V_4/\lambda^2_{\rm min}\ell^4\right]$ for some number $c$. The ratio of path integrals in the second line of \eqref{ratioPI} is thus the ratio of two exponentials of the same form, but with a priori two different coefficients $c$:
\ba
\frac{\int{\cal D}({\rm bosons})_E~\left|{\rm det}[H+im_5]\right| e^{-S_E}}{\left|\int{\cal D}({\rm bosons})_E~{\rm det}[H+im_5]e^{-S_E}\right|}
=\frac{e^{V_4/\ell^4(c m_5^2/\lambda^2_{\rm min}+\cdots)}}{e^{V_4/\ell^4(c' m_5^2/\lambda^2_{\rm min}+\cdots)}}.
\ea
Hence a priori it may depart exponentially from unity. Unless we are able to prove otherwise, no bound on \eqref{ratioPI} follows.

We are confident that some generalization of \cite{Vafa:1983tf} should be possible, but we were not able to find it yet. We leave this task to future work.

\subsection{Chiral Lagrangian: a consistency check}
\label{sec:chiralLag}

According to Sections \ref{sec:Pconservation} and \ref{sec:Flavorconservation} we expect that vector-like symmetries and parity remain conserved in all our models with a positive fermionic determinant and P-even, flavor-singlet scalars. We here provide supporting evidence of these results by showing that our conclusions are consistent with a chiral Lagrangian description.

For definiteness consider again the toy model \eqref{pseudoL}, where there is a gauge- and flavor-singlet P-even scalar and $N_f=1$. When $M_\phi=0$ the theory possesses a $SU(2)\times Z_{4T_{\cal R}/K}$ flavor symmetry, where $Z_{4T_{\cal R}/K}$ is the non-anomalous discrete subgroup of the axial $U(1)_A$. Here $2\pi K$ is the periodicity of the topological angle and empirically we find that for all real representations $T_{\cal R}=nK$ (for $n$ integer) \cite{Aharony:2013hda}. The symmetry is explicitly broken to $U(1)_B\times Z_{2}$ by $M_\phi\neq0$, where the baryon number $U(1)_B\subset SU(2)$ has been discussed before \eqref{pseudoL} whereas $Z_2\in SU(2)\times Z_{4T_{\cal R}/K}$ acts as $\psi_\pm\to \pm i\psi_\mp$.

We are interested in studying the effective field theory below the confinement scale $\Lambda$ of the non-abelian dynamics in the limit of small $|m|/\Lambda$ and small $|y|$. With $y=0$ the long distance physics is described by a non-linear sigma model associated to the symmetry breaking pattern $SU(2)\to U(1)_B$, as it follows from \cite{Vafa:1984xg}. The relevant low energy degrees of freedom are two Nambu-Goldstone modes, collected in a 2 by 2 matrix $\Sigma=\xi\sigma_1\xi^t$ with $\xi$ special unitary, as well as (at sufficiently large number of ``colors" $N_c$) the $\eta'$ meson. The Goldstone modes describe a complex scalar with $U(1)_B$ charge $q_B=2$.  In addition we allow the possibility that the mass of $\phi$ is arbitrary small and include also this degree of freedom. At the first non-trivial order in $|m|/\Lambda$ and $|y|$ the scalar potential reads~\footnote{The parameters $m,y$ here are those renormalized at the scale $\Lambda$ and not the bare ones, as in the rest of the paper.}
\ba\label{chiralPOT}
V_{m,\phi}
=V_{0,0}(\theta+2T_{\cal R}\eta')+\left[(am+by\phi)\Lambda^3{\rm Tr}\left[\sigma_1 \Sigma\right]e^{i\eta'}+{\rm hc}\right]+V_\phi.
\ea
Here $V_{0,0}(\theta+2T_{\cal R}\eta')$ is the main potential for $\eta'$ \cite{Witten:1979vv}, $V_\phi$ only depends on $\phi$, and $a,b$ are unknown numbers. We assume parity in the UV and so $\theta=0$ with $m,y$ real. Furthermore, because P is not broken by physics at zeroth order in $y$, we take $a$, $b$ to be real.

The derivation of the explicit vacuum configuration at first non-trivial order in $|m|/\Lambda$, $y$ is rather straightforward. The vacuum expectation value $\langle\eta'\rangle$ is obtained by minimizing the unknown, periodic function $V_{0,0}(2T_{\cal R}\eta')$ and is insensitive to the small perturbation ${\cal O}(|m|/\Lambda, y)$. Hence its value must be P-conserving, namely $\langle\eta'\rangle=0\sim\pi$. The perturbation introduces a tadpole for $\phi$ and lifts the flat direction parametrized by the Nambu-Goldstone matrix. The tadpole destabilizes the scalar by an amount that depends crucially on $V_\phi$. We assume $V_\phi$ has some non-trivial curvature, i.e we avoid fine-tuned scenarios in which $V_\phi$ is flat. In this case $y$ may be chosen sufficiently small so that the effective ``mass" $y\langle\phi\rangle$ is still within the regime of validity of the chiral Lagrangian analysis. Under this hypothesis the Nambu-Goldstone vacuum configuration can be shown to be 
\ba
\langle\Sigma\rangle ={\rm sign}(am+by\langle\phi\rangle)e^{-i\langle\eta'\rangle}~\sigma_1. 
\ea
Consistently with our earlier results, the vacuum solution conserves P and $U(1)_B\times Z_2$ for any value $|m|,|y\langle\phi\rangle|\ll\Lambda$. This remains true whatever the sign of the ${\cal O}(m,y)$ perturbation is. In fact, within the chiral Lagrangian there exists a non-anomalous transformation $\eta'\to\eta'+\pi$ (i.e. $\Sigma\to-\Sigma$) that changes  $(m,y\phi)\to(-m,-y\phi)$ (see the discussion before Section \ref{sec:gencharged} for the corresponding transformation of the fundamental fields). This is to be contrasted with QCD with $N_f$ odd, where a negative quark mass can flip the sign of the fermionic path integral and invalidate positivity --- a problem that would manifest itself at low-energy via tachyonic masses for some P-odd mesons which eventually trigger the breakdown of the vector-like flavor symmetries and P. In the gauge theories studied here, however, the Euclidean path integral has always positive sign and the chiral Lagrangian does indeed conform to this.

We stress that the chiral Lagrangian analysis we just presented cannot be viewed as a proof that $U(1)_B\times Z_2$ remains unbroken since we assumed a small Yukawa and made a perturbative expansion around $y=0$. At $y=0$, $m\neq0$ the theory conserves P and the Nambu-Goldstone bosons sit on the bottom of a $m$-dependent well, so small perturbations in $y$ cannot destabilize them. In the language of Section \ref{sec:Pconservation} this statement is also obvious. If one was allowed to perform a perturbative expansion around $y=0$ then one would find no zero-modes in the spectrum of $H$ with $m\neq0$; that expansion implies conservation of P. Yet, we see no reason in general for the Yukawa to be small, especially in view of the fact that $y$ is typically a relevant interaction. For this reason the analysis of this section is just a consistency check, rather than a proof.

Nevertheless, the chiral Lagrangian can teach us something non-trivial about the finite density behavior of these theories. At finite baryonic density the UV Lagrangian should be deformed by a real $U(1)_B$ chemical potential $\mu_{\rm c.p.}$:
\ba
{\cal L}\supset\mu_{\rm c.p.}\overline{\Psi}\gamma_0\Psi.
\ea
The operator $\gamma_5({\slashed{D}}+M_\phi+\mu_{\rm c.p.}\gamma_0)$ would not be hermitian anymore, our positivity argument would not apply, and $U(1)_B$-violating bilinears may form. Indeed, ref.~\cite{Kogut:1999iv} analyses the chiral Lagrangian of adjoint-QCD at finite density and finds that $\mu_{\rm c.p.}$ induces a potential for the Nambu-Goldstone bosons which, for sufficiently large values, triggers $U(1)_B$ breaking. (Note that the dominant correction in the Goldstone potential occurs at ${\cal O}(\mu_{\rm c.p.}^2)$ because the chemical potential is a CT-odd spurion.) We find that a qualitatively similar pattern occurs in our models with fundamental scalars.

\subsection{Mass inequalities}
\label{sec:massineq}

Our positivity theorem allows us to derive mass inequalities similar to those discussed in \cite{Weingarten:1983uj} for massive QCD. For concreteness we focus on a specific inequality in the simplest model \eqref{pseudoL}, but more may be derived along the same lines in this and other models.

Inequalities between two-point functions of operators can be straightforwardly translated into inequalities between the masses of narrow hadrons interpolated by such operators as long as there is no disconnected contribution. The latter was avoided in \cite{Weingarten:1983uj} by considering operators with off-diagonal flavor indices, which break the vector group and have vanishing expectation values in QCD. Here, we employ the fact that the baryon number remains conserved (see Section \ref{sec:Flavorconservation}) and consider hadronic fields of $U(1)_B$-charge $q_B=2$ 
\ba
{\mathbb M}_\Gamma(x)=\overline{\Psi}(x)\Gamma(\eta {\cal T})\overline{\Psi}^t(x),
\ea
with $\Gamma$ a combination of gamma matrices satisfying $\Gamma^t={\cal T}\,\Gamma\,{\cal T}^\dagger$ (and for later convenience normalized such that $\Gamma^\dagger\Gamma=1$). In an arbitrary bosonic background we get 
\ba
\langle\Omega|{\mathbb M}_\Gamma(x)\overline{\mathbb M}_\Gamma(y)|\Omega\rangle_A
&\equiv&\langle\Omega|\overline{\Psi}(x)\Gamma(\eta {\cal T})\overline{\Psi}^t(x)~\Psi^t(y)(\eta{\cal T})^\dagger\Gamma\Psi(y)|\Omega\rangle_A\\\no
&=&2\,{\rm tr}\left[S(y,x)\Gamma(\eta{\cal T}) S^t(y,x)(\eta{\cal T})^\dagger\Gamma\right]\\\no
&=&2\,{\rm tr}\left[S(y,x)\Gamma S^\dagger(y,x)\Gamma\right],
\ea
where $S(x,y)=\langle x|({{\slashed{D}}+M_\phi})^{-1}|y\rangle$ and we used $(\eta{\cal T}) S^t(x,y)(\eta{\cal T})^\dagger=S^\dagger(x,y)$. In these expressions the transpose and hermitian conjugation act on the Lorentz and gauge indices; likewise for the trace. The Cauchy-Schwartz inequality (here $\Gamma^\dagger\Gamma=1$ is employed) then leads to
\ba\label{ineq1}
\left|\langle\Omega|{\mathbb M}_\Gamma(x)\overline{\mathbb M}_\Gamma(y)|\Omega\rangle_A\right|\leq2\,{\rm tr}\left[S(y,x) S^\dagger(y,x)\right]=\langle\Omega|{\mathbb M}_{\Gamma=1}(x)\overline{\mathbb M}_{\Gamma=1}(y)|\Omega\rangle_A.
\ea
Positivity of the Euclidean path integral measure guarantees that the same inequality is satisfied even after averaging over the bosonic configurations. Hence, the absolute value of the Euclidean two-point function of ${\mathbb M}_\Gamma$ is maximized by the one of the P-even, Lorentz-singlet ${\mathbb M}_{\Gamma=1}$. Because by $U(1)_B$ conservation these hadron fields have vanishing vacuum expectation value, the above correlator has no disconnected piece. Inserting a complete set of 4-momentum eigenstates, and letting $|x-y|\to\infty$, eq. \eqref{ineq1} reveals that the lightest $q_B=2$ hadron is a P-even scalar. That result is somewhat expected physically, as that state morally corresponds to the would-be Nambu-Goldstone boson delivered by the approximate symmetry breaking pattern $SU(2)\to U(1)_B$ (see Section \eqref{sec:chiralLag}). The analogy is only suggestive, though, as our results apply to scenarios with an arbitrary real function $M_\phi$.

\section{Discussion}

This work presented new non-perturbative results in non-supersymmetric four-dimensional gauge theories with fundamental scalars. We showed that in a large class of P-invariant theories with Dirac flavors in the {\emph{real}} representation of the gauge group the Euclidean path integral is positive even when (relevant) Yukawa interactions to (charged or neutral) fundamental scalars are included. 

We discussed in detail how positivity of the measure can be employed to prove the conservation of parity. The proof of \cite{Vafa:1984xg} can be straightforwardly applied to argue that purely bosonic operators cannot be order parameters for P violation. The analysis of operators involving fermionic fields is more subtle. We demonstrated that P remains exactly conserved if and only if the spectrum of the fermionic operator ${\slashed{D}}+M_\phi$ admits no zero-modes (of finite bosonic measure). Even though we were not able to directly prove the absence of zero-modes, when combining our earlier results with the physically motivated assumption that any condensate that can develop indeed forms unless protected by an unbroken symmetry we obtain a fully-fledged proof of P conservation. We emphasize the key role played by such ``totalitarian principle" because it is often tacitly assumed in the literature; for example it is implicit in the proof of \cite{Vafa:1983tf} of the conservation of vector-like symmetries in massive QCD.

The absence of zero-modes was then central in our proof of the conservation of (continuous as well as discrete) flavor symmetries preserved by the mass and Yukawa couplings. Concretely, we showed that order parameters made of fermionic fields must have vanishing expectation value. The ``totalitarian principle" is invoked also here to establish that the same conclusion must apply to order parameters involving bosonic fields as well. We also attempted, without success, a generalizaton of an argument proposed in \cite{Vafa:1983tf} that constrains the long-distance behavior of the two-point functions of the order parameters and/or currents, which would confirm our claim that continuous symmetries cannot be spontaneously broken. We plan to come back to this consistency check in the future.

Finally, following \cite{Weingarten:1983uj}, positivity of the Euclidean path integral was employed to show that the lightest hadron with 2 units of $U(1)_B$ charge is a P-even scalar. 

It would be interesting to explore further the phenomenology of our scenarios, such as adjoint-QCD coupled to adjoint scalars, or $SO(N_c)$-QCD with scalars in the 2-index symmetric representation, as well as to look for possible generalizations, and eventually to explicitly test our claims via a numerical investigation on the lattice.

\section*{Acknowledgments}

We are grateful to L-X. Xu for collaborating in the early stages of the project and for pointing out a mistake in our original proof of P-conservation. LV would also like to thank D. Cassani, A. Wulzer, and D. Schaich for discussions. This research was partly supported by the Italian MIUR under contract 2017FMJFMW (PRIN2017), the “iniziativa specifica” Physics at the Energy, Intensity, and Astroparticle Frontiers (APINE) of Istituto Nazionale di Fisica Nucleare (INFN), and the European Union’s Horizon 2020 research and innovation programme under the Marie Sklodowska-Curie grant agreement No 860881-HIDDeN.


\appendix

\section{Parity non-violation in massive QCD}
\label{sec:masslessQCD}

In this appendix we discuss parity conservation in massive QCD. In particular we want to show that $\langle\overline{\Psi}i\gamma_5\Psi\rangle=0$ when $m>0$. An analogous discussion is presented in \cite{Einhorn:2002rm}.

To calculate the vacuum expectation value of the quark bilinear we follow the same procedure adopted in the rest of the paper: we perturb the QCD Lagrangian with $\delta{\cal L}=J\overline{\Psi}i\gamma_5\Psi$ and compute the path integral. Using again the basis $\left\{u_{\mu>0},\gamma_5u_{\mu>0},u_0^\pm\right\}$ we find 
\ba\label{detQCD}
{\slashed{D}}+m+iJ\gamma_5=\left(
\begin{matrix}
i\mu+m &  iJ & 0 &0\\
 iJ & -i\mu+m & 0 & 0\\
0 & 0 & m+iJ & 0\\
0 & 0 & 0 & m-iJ
\end{matrix}\right)
\ea
and
\ba\label{detQCD0}
{\rm det}[{\slashed{D}}+m+iJ\gamma_5]=\prod_{\mu>0}(\mu^2+m^2+J^2)(m+iJ)^{n_+}(m-iJ)^{n_-},
\ea
with $n_\pm$ denoting the numbers of zero-modes of ${\slashed{D}}$ with positive and negative chirality. Note that in QCD, as opposed to our scenarios, in general ${\rm det}[{\slashed{D}}+m+iJ\gamma_5]\neq{\rm det}[\gamma_5({\slashed{D}}+m+iJ\gamma_5)]$. In fact, as we saw in Section \ref{sec:posi}, ${\rm det}[\gamma_5]=(-1)^{N_\pm}(-1)^{n_-}$ and in QCD there is no reason for $N_\pm,n_-$ to be even unless $N_f$ is even as well.

The operator ${\slashed{D}}+m$ has a gapped spectrum because the quark mass $m$ behaves as IR-regulator. From the above expression we see that for any $m\neq0$ the limit ${J\to0}$ is smooth, at any volume. We can thus expand the determinant as follows
\ba
\ln {\rm det}[{\slashed{D}}+m+iJ\gamma_5]
&=&\ln {\rm det}[{\slashed{D}}+m]+i\frac{J}{m}(n_+-n_-)+{\cal O}(J^2)\\\no
&=&\ln {\rm det}[{\slashed{D}}+m]+i\frac{J}{m}\int{\rm d}^4x\,2T_{\cal R}\frac{g^2}{32\pi^2}F^A\widetilde F^A+{\cal O}(J^2V_4).
\ea
where according to the index theorem we re-wrote $n_+-n_-=2T_{\cal R}\nu$ in terms of the topological charge. (In QCD $2T_{\cal R}=N_f$.) At this point the partition function is analogous to \eqref{EuclideanPI} and the argument following \eqref{El>E0} can be used to argue that $\langle\overline{\Psi}i\gamma_5\Psi\rangle=0$ for any $m>0$.

\end{document}